\documentclass[proof]{pasj00}
\SetRunningHead{Y. Hasegawa and T. Tsuribe}{Kelvin-Helmholtz Instabilities in Multiple-Sized Dust Layers}
\Received{2013/04/25}
\begin{document}
\title{Kelvin-Helmholtz Instabilities in Multi-Sized Dust Layers}
\author{Yukihiko \textsc{Hasegawa}, and Toru \textsc{Tsuribe}}
\affil{Department of Earth and Space Science, Graduate School of Science, Osaka University, Toyonaka, Osaka 560-0043}
\email{hasegawa@vega.ess.sci.osaka-u.ac.jp, tsuribe@vega.ess.sci.osaka-u.ac.jp}
\maketitle

\begin{abstract}

We examine the effect of the dust size distribution on Kelvin-Helmholtz instabilities in the protoplanetary disk with dust sedimentation. With newly taking into account the dust size distribution, the growth rate of the Kelvin-Helmholtz instability is calculated using the linear stability analysis with the dust density distribution consistent with sedimentation. Dust abundance required for gravitational instabilities before the Kelvin-Helmholtz instability is derived from the linear stability analysis, and it is found that the required dust abundance significantly coincides with that estimated from the Richardson number. It is also found that when the dust size distribution is taken into account, the critical Richardson number for the onset of the Kelvin-Helmholtz instability tends to increase with dust abundance. This result is different from that in the case without the dust size distribution.

\end{abstract}

\section{Introduction}

In the process of the planetesimal formation, motion and growth of dust aggregates in the protoplanetary disks are key issues. The significant problem in the process of the planetesimal formation is the radial drift of meter-sized dust (\cite{key-A76}). To resolve this problem, gravitational instabilities (GI) in a thin dust layer are considered as one of the solutions (\cite{key-G73}; \cite{key-S83}). GI is expected to produce km-sized aggregates in short dynamical time scale when the density of the dust layer exceeds the critical density that is given by
\begin{equation}
\rho _\mathrm{c} = 3.6 \times 10^{-7} \left( {r} / {1 [\mathrm{AU}]} \right)^{-3} \, [\mathrm{g} \, \mathrm{cm}^{-3}] \mathrm{,} \label{eq:01}
\end{equation}
where $r$ is the heliocentric distance (\cite{key-S83}). The thin dust layer is supposed to form from dust aggregates settling toward the midplane (\cite{key-N81}, 1986). However, vertical gradient of dust abundance is indicated to induce the vertical shear of the azimuthal velocity (\cite{key-C93}; \cite{key-S98}), and the shear is expected to induce the Kelvin-Helmholtz instability (KHI) (\cite{key-C61}). KHI is expected to induce shear-driven turbulence that prevents dust aggregates from settling for GI (\cite{key-C93}; \cite{key-S98}; \cite{key-S01}; \cite{key-M06}). Previous studies indicate that GI tends to be occur when dust abundance is high (\cite{key-S98}; \cite{key-J07}). However, with taking account of dust growth, Hasegawa \& Tsuribe (2013, hereafter HT13) indicated that KHI tended to occur before GI even in the case with large dust abundance. In HT13, in order to investigate dust densities in the midplane at the onset of KHI, we used the Richardson number (\cite{key-C61}) and a criterion that the minimum value of the Richardson number distribution $J_\mathrm{min}$ is the critical value $J_\mathrm{c} = 0.25$. However, $J_\mathrm{c} = 0.25$ is derived only for the case with the incompressible, inviscid and laminar fluid. There are many previous studies of the Richardson criterion and the critical Richardson number (\cite{key-S00}; \cite{key-G04}; \cite{key-G05}; \cite{key-C08}; \cite{key-B09}; \cite{key-L10}). They show that the critical Richardson number depends on the assumed density profile, and it is suggested that dust abundance and dust sedimentation are important for understanding KHI. Other than the critical Richardson number, \citet{key-S00} and \citet{key-M06} calculated the growth rate of KHI by solving the linear perturbation equation. However, dust densities used in \citet{key-S00} were limited to be smaller than the critical density for GI. \citet{key-M06} investigated only the case with a particular dust density distribution.

In this paper, the growth rate of KHI is calculated using the linear stability analysis with the dust density distribution that is consistent with their sedimentation in the protoplanetary disk. In \S 2, the models of gas and dust components in the protoplanetary disk in this paper and equations for the linear stability analysis are described. In \S 3, we show results for the growth rate and the critical Richardson number, and we discuss the validity of a criterion $J_\mathrm{min} = J_\mathrm{c} = 0.25$ used in HT13. In \S 4, we summarize our results.

\section{Models and Equations}

To calculate the growth rate of KHI, we use the dust density distribution that is consistent with their sedimentation. For simplicity, we concentrate on KHI induced by vertical dust sedimentation near the midplane. We do not take into account the inward orbital drifts of dust aggregates. We restrict ourselves to the case with $r = 1$ AU. The gas surface density $\Sigma _\mathrm{g}$ and the dust surface density $\Sigma _\mathrm{d}$ are assumed to be
\begin{equation}
\Sigma _\mathrm{g} = 1.7 \times 10^3 \, [\mathrm{g} \, \mathrm{cm}^{-2}] \mathrm{,} \label{eq:02}
\end{equation}
\begin{equation}
\Sigma _\mathrm{d} = 7.1 f_\mathrm{d} \, [\mathrm{g} \, \mathrm{cm}^{-2}] \mathrm{,} \label{eq:03}
\end{equation}
where $f_\mathrm{d}$ is the parameter for dust abundance, and $f_\mathrm{d} = 1$ corresponds to MMSN model (\cite{key-H81}). We assume that the dust layer is geometrically thin, and that the gas density is uniform in the dust layer and is assumed as
\begin{equation}
\rho _\mathrm{g} = {\Sigma _\mathrm{g}} / ({\sqrt{\pi } H_\mathrm{g}}) = 1.4 \times 10^{-9} \, [\mathrm{g} \, \mathrm{cm}^{-3}] \mathrm{,} \label{eq:04}
\end{equation}
where $H_\mathrm{g} = 4.7 \times 10^{-2} \mathrm{AU}$ is the scale height of the protoplanetary disk.

We assume that dust aggregates are small enough to couple strongly to gas. Thus, the mixture of gas and dust is treated as one component fluid. We calculate the Richardson number (\cite{key-C61}) as an indicator of KHI. The Richardson number is given by
\begin{equation}
J = - ({{\Omega _\mathrm{K}}^2 z} / {\rho _\mathrm{0}}) [{\partial \rho _\mathrm{d0} (z)} / {\partial z}] \left( {\partial v_0 } / {\partial z} \right)^{-2} \mathrm{,} \label{eq:05}
\end{equation}
where $z$ is the height from the midplane, $\Omega _\mathrm{K}$ is the Keplerian angular velocity, $\rho _\mathrm{d0} (z)$ is the unperturbed dust density, $\rho _\mathrm{0} \equiv \rho _\mathrm{g} + \rho _\mathrm{d0} (z)$, and $v_0$ is the unperturbed azimuthal velocity of the mixed fluid of gas and dust. In this paper, subscript 0 refers the unperturbed quantities, and subscript 1 refers the perturbed quantities. The unperturbed azimuthal velocity is given by the equilibrium of three forces, the gravity of the central star, the centrifugal force and the gas pressure gradient force, as
\begin{equation}
v_0 = \left( 1 - {\rho _\mathrm{g}} / {\rho _\mathrm{0}} \right) \eta v_\mathrm{K} \mathrm{,} \label{eq:06}
\end{equation}
where $v_\mathrm{K}$ is the circular Keplerian velocity, and $\eta = - ({H_\mathrm{g}}^2 / 4 r^2) (\partial \ln P_\mathrm{g} / \partial \ln r) = 1.8 \times 10^{-3}$, where $P_\mathrm{g}$ is the gas pressure. In this paper, we adopt the local Cartesian coordinate system rotating around the central star at $r = 1$ AU with the azimuthal velocity $(1 - \eta ) v_\mathrm{K}$ (\cite{key-S00}). From equations (\ref{eq:05}) and (\ref{eq:06}), we obtain
\begin{equation}
J = - [{z} / ({\eta ^2 r^2})] ({{\rho _\mathrm{0}}^3} / {{\rho _\mathrm{g}}^2}) \left[ {\partial \rho _\mathrm{d0} (z)} / {\partial z} \right] ^{-1} \mathrm{.} \label{eq:07}
\end{equation}
\citet{key-C61} showed that KHI is expected to be induced when $J < J_\mathrm{c} = 0.25$.

We adopt the initial dust density distribution that is the same as \citet{key-S01}, and that is given by
\begin{eqnarray}
\rho _\mathrm{d0} (z) = \left\{ \begin{array}{ll}
({\rho _\mathrm{ini}} / {2}) \left[ 1 + \cos \left( \pi {z} / {z_\mathrm{d}} \right) \right] & (|z| < z_\mathrm{d}) \\
0 & (|z| \ge z_\mathrm{d}) \\
\end{array} \right. \mathrm{,} \label{eq:08}
\end{eqnarray}
where $z_\mathrm{d}$ is the time dependent scale height of the dust density profile with $z_\mathrm{d} = H_\mathrm{g}$ at the initial state, and $\rho _\mathrm{ini}$ is the initial dust density at the midplane and is given by
\begin{equation}
\rho _\mathrm{ini} = {\Sigma _\mathrm{d}} / {H_\mathrm{g}} \mathrm{.} \label{eq:09}
\end{equation}
In this paper, we investigate the stability for the series of the density distribution induced by sedimentation. Actually, the stability analysis with the time-dependent evolution of the dust density can not be reduced to the eigenvalue problem. However, in this paper, density distributions for each instant of time are approximated by steady distributions since dust sedimentation is slow enough. In such a case, the dust density profile evolves in a self-similar manner  with a single size dust without growth (\cite{key-G04}). On the other hand, in the case with the dust size distribution, the dust density profile is not self-similar during dust sedimentation (HT13). In this paper, we consider the dust size distribution but neglect dust growth for simplicity. We consider $N_\mathrm{d}$ kinds of sizes (radii) for dust aggregates. In this case, the dust density distribution $\rho _\mathrm{d0} (z)$ is given by
\begin{equation}
\rho _\mathrm{d0} (z) = \sum _{n = 1}^{N_\mathrm{d}} \rho _{s_n} (z) \mathrm{,} \label{eq:10}
\end{equation}
where $\rho _{s_n} (z)$ is the density of dust component with size $s_n$ at $z$. For simplicity, we assume that the initial size distribution is given by
\begin{equation}
\rho _{s_n} (z) = \rho _{s_{\mathrm{min}}} (z) = \rho _\mathrm{d0} (z) / N_\mathrm{d} \mathrm{,} \label{eq:11}
\end{equation}
where $s_{\mathrm{min}}$ is the minimum size of dust aggregates. In this case, $\rho _{s_n} (z)$ is given by
\begin{eqnarray}
\rho _{s_n} (z) = \left\{ \begin{array}{ll}
\frac{\rho _\mathrm{ini}}{2 N_\mathrm{d}} \frac{H_\mathrm{g}}{z_{s_n}} \left[ 1 + \cos \left( \frac{\pi z}{z_{s_n}} \right) \right] & (|z| < z_{s_n}) \\
0 & (|z| \ge z_{s_n}) \\
\end{array} \right. \mathrm{,} \label{eq:12}
\end{eqnarray}
where $z_{s_n}$ is the scale height of the dust density profile for dust with size $s_n$, and $z_{s_n}$ is derived in HT13 as
\begin{equation}
{z_{s_n}} / {H_\mathrm{g}} = \left( {z_{s_{\mathrm{min}}}} / {H_\mathrm{g}} \right)^{s_n / s_{\mathrm{min}}} \mathrm{,} \label{eq:13}
\end{equation}
where $z_{s_\mathrm{min}}$ is the scale height of the dust density profile for minimum dust aggregates and $z_{s_\mathrm{min}}$ determines $z_\mathrm{d}$. We assume that $N_\mathrm{d} = 11$ and that $s_n$ is given by $s_{\mathrm{min}}$, $1.1 s_{\mathrm{min}}$, $1.2 s_{\mathrm{min}}$, $\dots$, and $2 s_{\mathrm{min}}$. If the values of $f_\mathrm{d}$ and $z_{s_{\mathrm{min}}}$ are given, the dust density distribution $\rho _\mathrm{d0} (z)$ is obtained from equations (\ref{eq:03}), (\ref{eq:09}), (\ref{eq:10}), (\ref{eq:12}) and (\ref{eq:13}).

We calculate the growth rate of KHI for the case with the unperturbed dust density described above. To calculate the growth rate, we solve the linear perturbation equation derived by \citet{key-S00}. Perturbed quantities are assumed to have the form as $F_1 = \hat{F}_1 (z) \exp [i (k y - \omega t)]$, where $k$ is the azimuthal wave number, $y$ is the azimuthal coordinate , and $\omega = \omega _r + i \omega _i$ is the complex angular frequency. For perturbed quantities, we restrict ourselves to the case with the radial wave number to be zero because this is the most unstable mode. Hereafter $\hat{~}$ is omitted, and the equation is given by
\begin{eqnarray}
& & \frac{d^2 w_1}{dz^2} + \frac{1}{\rho _\mathrm{0}} \frac{d \rho _\mathrm{0}}{dz} \frac{dw_1}{dz} - \left( k^2 + \frac{1}{\bar{v}} \frac{d^2 v_0}{dz^2} + \frac{1}{\rho _\mathrm{0}} \frac{d \rho _\mathrm{0}}{dz} \frac{1}{\bar{v}} \frac{dv_0}{dz} \right. \nonumber \\
&& \left. {~~~~~~~~~~~~~~~~~~~~~~~~~~~~~~} + \frac{{\Omega _\mathrm{K}}^2 z}{{\bar{v}}^2} \frac{1}{\rho _\mathrm{0}} \frac{d \rho _\mathrm{0}}{dz} \right) w_1 = 0 \mathrm{,} \label{eq:14}
\end{eqnarray}
where $w_1$ is the perturbed vertical velocity, and $\bar{v} \equiv v_0 - \omega / k$. Boundary conditions are given by
\begin{equation}
w_1 = 0 {~} \mathrm{at} {~} z = 0 {~} (\mathrm{odd {~} mode}) \mathrm{,} \label{eq:15}
\end{equation}
\begin{equation}
{dw_1} / {dz} = 0 {~} \mathrm{at} {~} z = 0 {~} (\mathrm{even {~} mode}) \mathrm{,} {~} \mathrm{and} \label{eq:16}
\end{equation}
\begin{equation}
{dw_1} / {dz} + \left[ k - ({1} / {\bar{v}}) ({dv_0} / {dz}) \right] w_1 = 0 {~} \mathrm{at} {~} z = z_\mathrm{d} \mathrm{.} \label{eq:17}
\end{equation}
With assuming the dust abundance parameter $f_\mathrm{d}$, the azimuthal wave number $k$ and the dust density at the midplane $\rho _\mathrm{d0} (0)$, we solve the linear perturbation equation (\ref{eq:14}) numerically with the boundary condition (\ref{eq:17}) for the odd mode (\ref{eq:15}) and the even mode (\ref{eq:16}), respectively. At fixed dust abundance $f_\mathrm{d}$ and the dust density $\rho _\mathrm{d0} (0)$, we obtain eigenvalues, $\omega _r$ and $\omega _i$, as functions of the wave number $k$, respectively, and the maximum growth rate of KHI $\omega _{i, \mathrm{max}}$ is also obtained. At fixed dust abundance $f_\mathrm{d}$, the maximum growth rate $\omega _{i, \mathrm{max}}$ is a function of the dust density. Especially, we regard the minimum dust density in the case with $\omega _{i, \mathrm{max}} \ge 0$ as the possible dust density attained without KHI. From the condition of $\omega _{i, \mathrm{max}} \ge 0$, the possible dust density at the midplane $\rho _\mathrm{d0} (0)$ is finally derived as a function of dust abundance $f_\mathrm{d}$.

In this paper, we compare the following two cases. The case with single-sized dust without the size distribution is labeled with A, and the case with multi-sized dust with the size distribution is labeled with B.

\section{Results and Discussion}

It is found that even modes are more unstable than odd modes except for two cases: case B with $f_\mathrm{d} = 1$ and $\rho _\mathrm{d0} (0)$ $< 0.33 \rho _\mathrm{g}$, and case B with $f_\mathrm{d} = 2$ and $\rho _\mathrm{d0} (0) < 0.61 \rho _\mathrm{g}$.

\begin{figure}
  \begin{center}
    \FigureFile(74mm,74mm){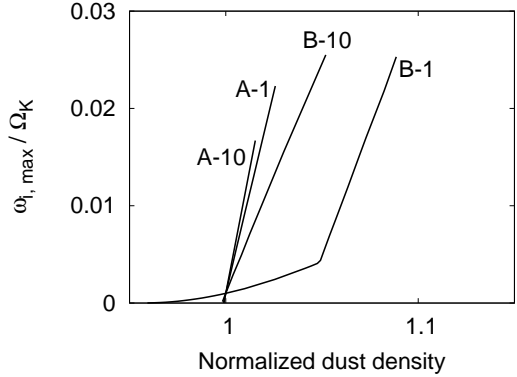}
  \end{center}
  \caption{Maximum growth rates of KHI $\omega _{i, \mathrm{max}}$ normalized by the Keplerian angular velocity $\Omega _\mathrm{K}$ as functions of the dust density at the midplane normalized by that in the case with $\omega _{i, \mathrm{max}} = 10^{-3} \Omega _\mathrm{K}$. A-1 means the case A with $f_\mathrm{d} = 1$, A-10 means the case A with $f_\mathrm{d} = 10$, B-1 means the case B with $f_\mathrm{d} = 1$, and B-10 means the case B with $f_\mathrm{d} = 10$.}\label{fig:01}
\end{figure}

Figure \ref{fig:01} shows maximum growth rates of KHI as functions of the dust density at the midplane. From Figure \ref{fig:01}, the crossing point of each line with $\omega _{i, \mathrm{max}} = 0$ is the possible dust density attained before KHI. As seen in Figure \ref{fig:01}, it is seen that the dust density in the case with $\omega _{i, \mathrm{max}} = 10^{-3} \Omega _\mathrm{K}$ approximates the dust density in the case with $\omega _{i, \mathrm{max}} = 0$. Since it is technically difficult to calculate the case with exactly $\omega _{i, \mathrm{max}} = 0$ especially in the case with the large dust abundance parameter $f_\mathrm{d}$, we use the dust density for $\omega _{i, \mathrm{max}} = 10^{-3} \Omega _\mathrm{K}$ instead of that for $\omega _{i, \mathrm{max}} = 0$ except for case B with $f_\mathrm{d} = 1$ and B with $f_\mathrm{d} = 2$. For the two cases B with $f_\mathrm{d} = 1$ and B with $f_\mathrm{d} = 2$, we use the dust density for $\omega _{i, \mathrm{max}} = 10^{-5} \Omega _\mathrm{K}$ instead of that for $\omega _{i, \mathrm{max}} = 0$.

\begin{figure}
  \begin{center}
    \FigureFile(74mm,74mm){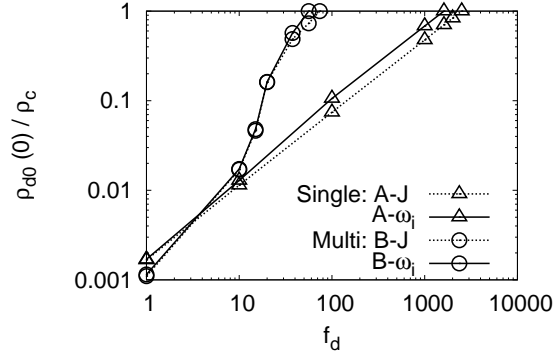}
  \end{center}
  \caption{Dust densities at the midplane $\rho _\mathrm{d0} (0)$ normalized by the critical density for GI $\rho _\mathrm{c}$ as functions of the dust abundance parameter $f_\mathrm{d}$ for the cases A (triangles) and B (circles). Dotted lines show the densities with the minimum value of the Richardson number distribution $J_\mathrm{min} = 0.25$. Solid lines show the densities with the maximum growth rate of KHI $\omega _{i, \mathrm{max}} = 10^{-5} \Omega _\mathrm{K}$ (case B-1) or $10^{-3} \Omega _\mathrm{K}$ (other cases).}\label{fig:02}
\end{figure}

Figure \ref{fig:02} shows dust densities at the midplane as functions of dust abundance. A label A-J refers the case A (with single-sized dust) with $J_\mathrm{min} = 0.25$, a label A-$\omega _{i}$ refers the case A with $\omega _{i, \mathrm{max}} = 10^{-3} \Omega _\mathrm{K}$, a label B-J refers the case B (with multi-sized dust) with $J_\mathrm{min} = 0.25$, and a label B-$\omega _{i}$ refers the case B with $\omega _{i, \mathrm{max}} = 10^{-5} \Omega _\mathrm{K}$ (case B-1) or $\omega _{i, \mathrm{max}} = 10^{-3} \Omega _\mathrm{K}$ (other cases). From Figure \ref{fig:02}, it is found that for both cases A and B the dust abundance parameters significantly coincide between the cases with $\omega _{i, \mathrm{max}} \sim 0$ and with $J_\mathrm{min} = 0.25$. Especially, the dust density at the midplane $\rho _\mathrm{d0} (0)$ in the case B-$\omega _{i}$ with $f_\mathrm{d} \le 20$ remarkably coincides with that in the case B-J, and the dotted and solid lines look like overlapping.

In Figure \ref{fig:02}, it is seen that dust abundance parameters required to achieve $\rho _\mathrm{d0} (0) / \rho _\mathrm{c} = 1$ for the case with $J_\mathrm{min} = 0.25$ are $f_\mathrm{d} = 2.5 \times 10^3$ in the case A-J and $f_\mathrm{d} = 74$ in the case B-J. On the other hand, in the case with $\omega _{i, \mathrm{max}} \sim 0$, it is seen that dust abundance parameters required for $\rho _\mathrm{d0} (0) / \rho _\mathrm{c} = 1$ are $f_\mathrm{d} = 1.6 \times 10^3$ in the case A-$\omega _{i}$ and $f_\mathrm{d} = 56$ in the case B-$\omega _{i}$. For above results, values of dust abundance parameters $f_\mathrm{d}$ are larger than values in HT13 that are $f_\mathrm{d} = 6.6 \times 10^2$ in the case A-J and $f_\mathrm{d} < 50$ in the case B-J. This is because the initial dust density distribution in this paper is different from the Gaussian profile used in HT13. From equation (\ref{eq:07}), the Richardson number distribution depends strongly on the dust density profile, so dust abundance required for GI varies with the initial dust density distribution even with the same $J_\mathrm{min}$.

From Figure \ref{fig:02}, dust abundance parameters required to achieve $\rho _\mathrm{d0} (0) / \rho _\mathrm{c} = 1$ in the case with multi-sized dust is much smaller than these in the case with single-sized dust. This tendency is similar to the result in HT13. It is suggested that this tendency is independent on the initial distribution of the dust density.

In summary, it is found that, for both cases with single-sized dust and with multi-sized dust, the dust abundance parameters significantly coincide between the cases with $\omega _{i, \mathrm{max}} \sim 0$ and with $J_\mathrm{min} = 0.25$. Thus, it is seen that dust abundance parameters that HT13 calculated using a criterion $J_\mathrm{min} = J_\mathrm{c} = 0.25$ are reasonable at least with errors of a factor of 2.

\begin{figure}
  \begin{center}
    \FigureFile(74mm,74mm){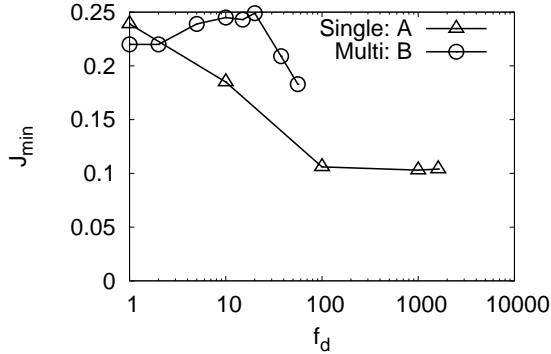}
  \end{center}
  \caption{Minimum Richardson numbers in the case with $\omega _{i, \mathrm{max}} = 10^{-5} \Omega _\mathrm{K}$ (cases B-1 and B-2) or $\omega _{i, \mathrm{max}} = 10^{-3} \Omega _\mathrm{K}$ (other cases) as functions of the dust abundance parameter $f_\mathrm{d}$ for the case A (triangles) and for the case B (circles).}\label{fig:03}
\end{figure}

For the case with fixed $f_\mathrm{d}$ and $\omega _{i, \mathrm{max}}$, we calculate the spatial profile of a Richardson number as a function of $z$. Then we obtain a minimum value of the Richardson number in this profile as the critical Richardson number. Figure \ref{fig:03} shows the minimum values of Richardson numbers in the case with $\omega _{i, \mathrm{max}} = 10^{-5} \Omega _\mathrm{K}$ (cases B-1 and B-2) or $\omega _{i, \mathrm{max}} = 10^{-3} \Omega _\mathrm{K}$ (other cases) as functions of dust abundance. For the case A (with single-sized dust), the critical Richardson number tends to decrease with increasing dust abundance. On the other hand, for the case B (with multi-sized dust), the critical Richardson number tends to increase with dust abundance in the case with $f_\mathrm{d} \le 20$ and to approach 0.25 in the case with $5 \le f_\mathrm{d} \le 20$. This result is different from that in the case A. From these results, in the case with the dust size distribution it is suggested that the critical Richardson number might be approximated by 0.25 and that $J_\mathrm{min} = 0.25$ condition is applicable, at least in the case with $5 \lesssim f_\mathrm{d} \lesssim 20$.

\section{Summary}

In this paper, we have calculated the growth rate of the Kelvin-Helmholtz instability using the dust density distribution that is consistent with dust sedimentation in the protoplanetary disk. We assumed the thin dust layer and the uniform gas density. Dust aggregates were assumed to be small, so the mixture of gas and dust aggregates was treated as one component fluid. We considered the dust size distribution without the dust growth. We have solved the linear perturbation equation to calculate the growth rate, and we have examined the dust abundance required for the gravitational instability of the dust layer.

Our study shows the following results: (1) For the dust abundance required for the gravitational instability, in the both cases without and with the dust size distribution, the dust abundance for the case with $\omega _{i, \mathrm{max}} \sim 0$ is about the same as that derived by the condition $J_\mathrm{min} = 0.25$. Thus, it is seen that dust abundance parameters that HT13 calculated using a criterion $J_\mathrm{min} = J_\mathrm{c} = 0.25$ are reasonable at least with errors of a factor of 2. (2) The critical Richardson number is affected by the dust size distribution. In the case without the dust size distribution, the critical Richardson number tends to decrease with increasing dust abundance. On the other hand, in the case with the dust size distribution, the critical Richardson number tends to increase and approach 0.25 with increasing dust abundance in the case with $5 \lesssim f_\mathrm{d} \lesssim 20$. Thus, it is indicated that $J_\mathrm{min} = 0.25$ condition is applicable in the case with the dust size distribution in the case with $5 \lesssim f_\mathrm{d} \lesssim 20$.

In this paper, we neglect the streaming instability (SI) (\cite{key-Y05}; \cite{key-J07}). \citet{key-Y05} showed that the growth rate of SI decreases with decreasing the stopping time of the dust aggregate. It is shown that dust aggregates are highly porous (\cite{key-W98}; \cite{key-K99}; \cite{key-O09}). Porous dust aggregates have small stopping times; thus KHI might occur before SI. This problem will be addressed in subsequent papers.

The Coriolis and tidal forces are not included in the linear perturbation equation (\ref{eq:14}). \citet{key-I03} performed the linear stability analysis including these forces and found that the tidal force plays an important role in the stabilization. However, as a first step to focus on the effect of the dust size distribution, in this paper we ignore the Coriolis and tidal forces for simplicity. In future work, we should take into account these forces.

\bigskip

\noindent We thank Fumio Takahara for fruitful discussion and continuous encouragement.

\end{document}